\documentclass[aps,prl,a4paper,twocolumn,showpacs,floatfix]{revtex4}
\usepackage[dvips]{epsfig}
\begin{document}
\title{Ultracold Li + Li$_2$ collisions: bosonic and fermionic cases}
\author{Marko T. Cvita\v{s}}
\author{Pavel Sold\'{a}n}
\author{Jeremy M. Hutson}
\affiliation{Department of Chemistry, University of Durham, South Road,
Durham, DH1~3LE, England}
\author{Pascal Honvault}
\author{Jean-Michel Launay}
\affiliation{UMR 6627 du CNRS, Laboratoire de Physique des Atomes,
Lasers, Mol\'ecules et Surfaces, Universit\'e de Rennes, France}

\date{\today}

\begin{abstract}
We have carried out quantum dynamical calculations of vibrational
quenching in Li + Li$_2$ collisions for both bosonic $^7$Li and
fermionic $^6$Li. These are the first ever such calculations
involving fermionic atoms. We find that for the low initial
vibrational states considered here ($v\le 3$), the quenching rates
are {\it not} suppressed for fermionic atoms. This contrasts with
the situation found experimentally for molecules formed via
Feshbach resonances in very high vibrational states.
\end{abstract}

\maketitle

\font\smallfont=cmr7

There is at present great interest in the properties of cold and
ultracold molecules. At the end of 2003, three groups
\cite{Grimm03a,Jin03a, Kett03a} created long-lived molecular
Bose-Einstein condensates (BECs) from ultracold atomic gases by
making use of magnetically tunable Feshbach resonances. The
molecules are formed in highly excited vibrational states, and any
vibrational relaxation will result in trap loss. A crucial
breakthrough was the use of fermionic isotopes of the atomic gases
($^{6}$Li, $^{40}$K) tuned to large positive scattering lengths;
it was found that in this case the atom-molecule and
molecule-molecule inelastic collisions are strongly suppressed
\cite{Jin03b,Sal03,Hulet03a,Grimm03b}, and the molecular cloud
consisting of weakly bound dimers of fermions exhibits a
remarkable stability against collisional decay. By contrast, for
weakly bound dimers of bosonic atoms the vibrational relaxation is
fast, so that the corresponding molecular cloud decays quickly and
quantum degeneracy can be achieved only transiently
\cite{Kett99,Wiem02,Grimm03c,Remp04,Kett04}. Petrov {\em et al.}
\cite{Gora03} have explained the important difference between
molecular gases of weakly bound dimers formed from bosonic and
fermionic atoms in terms of different quantum statistics for the
atoms.

In this Letter, we carry out the first full quantum dynamics
calculations for spin-polarized collisions of ultracold
homonuclear molecules for Li + Li$_2$ in both the bosonic ($^7$Li)
and fermionic ($^6$Li) cases. We find that for low-lying
vibrational states there is no systematic suppression of the
vibrational relaxation (quenching) rates for fermions, even when
the scattering length is large and positive. The analysis of
Petrov {\em et al.} \cite{Gora03} applies only when the scattering
length is large and positive {\it and} the molecules are in their
highest-lying vibrational state.

\begin{figure} [htbp]
\begin{center}
\rotatebox{270}{ \resizebox{7cm}{!}
{\includegraphics{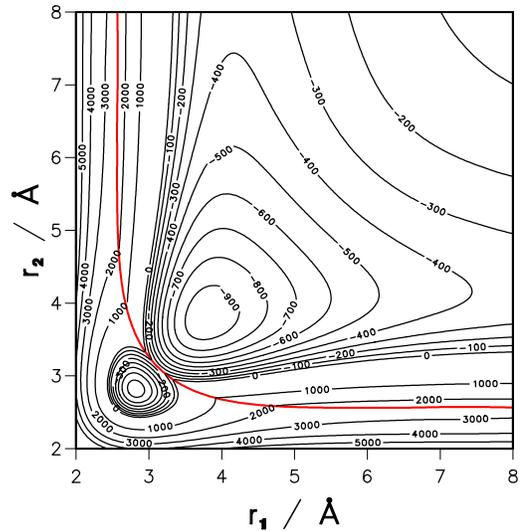}}} \caption{Cut through Li$_3$
quartet surface at collinear geometries, showing seam between
$^4\Sigma$ and $^4\Pi$ states. Contours are labelled in cm$^{-1}$.
The two minima are at depths of 760 and 950 cm$^{-1}$. }
\label{seam}
\end{center}
\end{figure}

Quantum scattering calculations require accurate potential energy
surfaces. We have previously shown that the interaction potential
for spin-polarized (quartet) Li + Li$_2$ is highly non-additive
\cite{Sol03}, with a well depth 4 times that of the sum of Li-Li
pair potentials. In the present work, we use a global Li$_3$
potential obtained from all-electron coupled-cluster electronic
structure calculations, which will be described in detail
elsewhere \cite{Cvi04pot}. In parallel work, Colavecchia {\em et
al.} \cite{Col03} have obtained another potential surface based on
pseudopotential calculations. However, we chose to use our own
surface in the present work because it takes better account of
some features of the potential. In particular, as pointed out in
Ref.\ \onlinecite{Sol03}, there is a seam of conical intersections
at collinear geometries, between the $^4\Sigma$ surface that is
the ground state at long range and a $^4\Pi$ surface that becomes
the ground state inside $r_1=r_2=3.1$ \AA. The resulting cusp
drops to energies just below the atom-diatom threshold, as shown
in Fig. \ref{seam}, so may play a significant role in the
dynamics. Our potential takes better account of this seam than
that of ref.\ \onlinecite{Col03}.

The methods we use to carry out quantum dynamics calculations on
systems of this type have been described in our previous work on
Na + Na$_2$ \cite{Sol02,Quem04} and elsewhere in the context of
thermal reactive scattering \cite{hon04}, so a brief summary will
suffice here. The potential energy surface is barrierless, so that
it is essential to take reactive (atom exchange) collisions into
account. The positions of the nuclei are described in
hyperspherical democratic coordinates. The configuration space is
divided into inner and outer regions, and the boundary between
them is placed at a distance such that couplings due to the
residual atom-diatom interaction can be neglected outside the
boundary. In the inner region (hyperradius $\rho \le 45\ a_0$ in
the present case), we obtain the wavefunction for nuclear motion
by propagating a set of coupled equations in a diabatic-by-sector
basis that is obtained by diagonalizing a fixed-$\rho$ reference
Hamiltonian in a basis set of pseudohyperspherical harmonics. In
the outer region, we use the Arthurs-Dalgarno formalism
\cite{Dalgarno}, which is based on Jacobi coordinates, and compute
by inwards integration regular and irregular solutions of a radial
Schr{\"o}dinger equation which includes the isotropic ($R^{-6}$)
part of the interaction. Matching between wavefunctions in the
inner and outer regions yields the scattering S-matrix.

The difference between the $^6$Li and $^7$Li cases is implemented
by selecting the pseudohyperspherical harmonic basis functions
included in the basis set. For both bosons and fermions, we
consider spin-stretched states of all three atoms involved (states
with $|M_F|=F=F_{\rm max}$). For such states the nuclear spin
wavefunction is symmetric with respect to any exchange of nuclei.
However, the electronic wavefunction for the quartet state is
antisymmetric with respect to exchange of nuclei. Thus, to satisfy
the Pauli Principle, the wavefunction for nuclear motion must be
antisymmetric with respect to exchange of bosonic nuclei (for
fermionic alkali metal atoms) and symmetric with respect to
exchange of fermionic nuclei (for bosonic alkali metal atoms).

Asymptotically, the hyperspherical functions correlate with
atom-diatom functions; the diatom functions for Hund's case (b),
which is appropriate for Li$_2\ (^3\Sigma_u^+$), are labelled with
a vibrational quantum number $v$ and a mechanical rotational
quantum number $n$; $n$ couples with the diatomic electron spin
$s=1$ to give a resultant $j$, but for Li$_2$ the splittings
between states of the same $n$ but different $j$ are very small
and are neglected here. In the $^3\Sigma_u^+$ state, only even $n$
is allowed for $^7$Li$_2$ and only odd $n$ for $^6$Li$_2$. In the
outer region, we include rovibrational states with $v=0,1,...,7$
with all even rotational levels up to $n_{\rm max}=32, 30, 28, 24,
22, 18, 14 ,10$ for bosonic $^7$Li$_2$ and rovibrational states
with $v=0,1,...,7$ with all odd rotational levels up to $n_{\rm
max}=31, 27, 25, 23, 19, 17, 13, 7$ for fermionic $^6$Li$_2$. In
the inner region, the number of coupled equations for bosonic
atoms varies from 97 for total angular momentum $J=0$ to 827 for
$J=10$ and for fermionic atoms it varies from 85 for $J=0^{+}$ to
782 for $J=11^{-}$.

\begin{figure} [htbp]
\begin{center}
\rotatebox{270}{ \resizebox{7cm}{!}
{\includegraphics{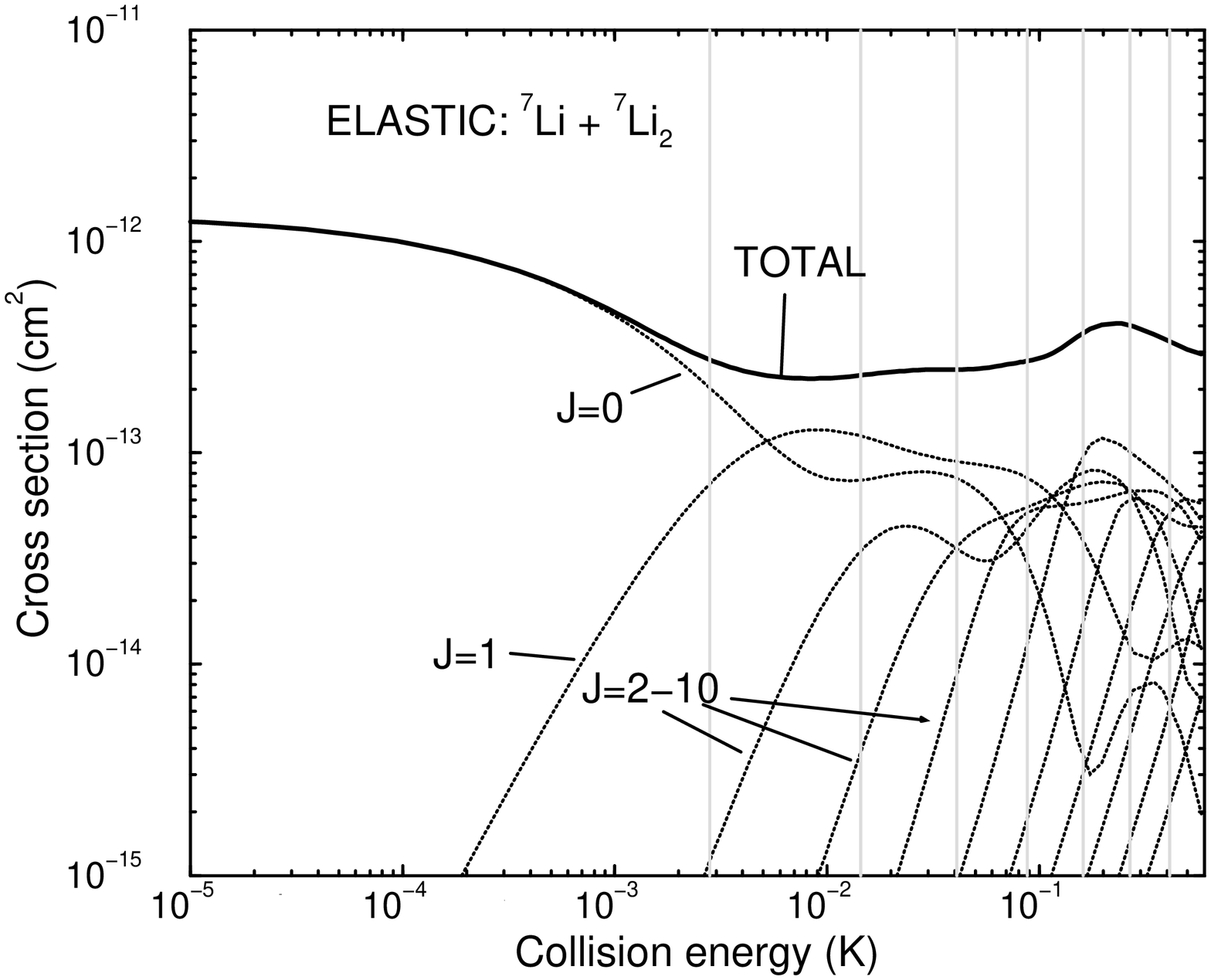}}} \rotatebox{270}{
\resizebox{7cm}{!} {\includegraphics{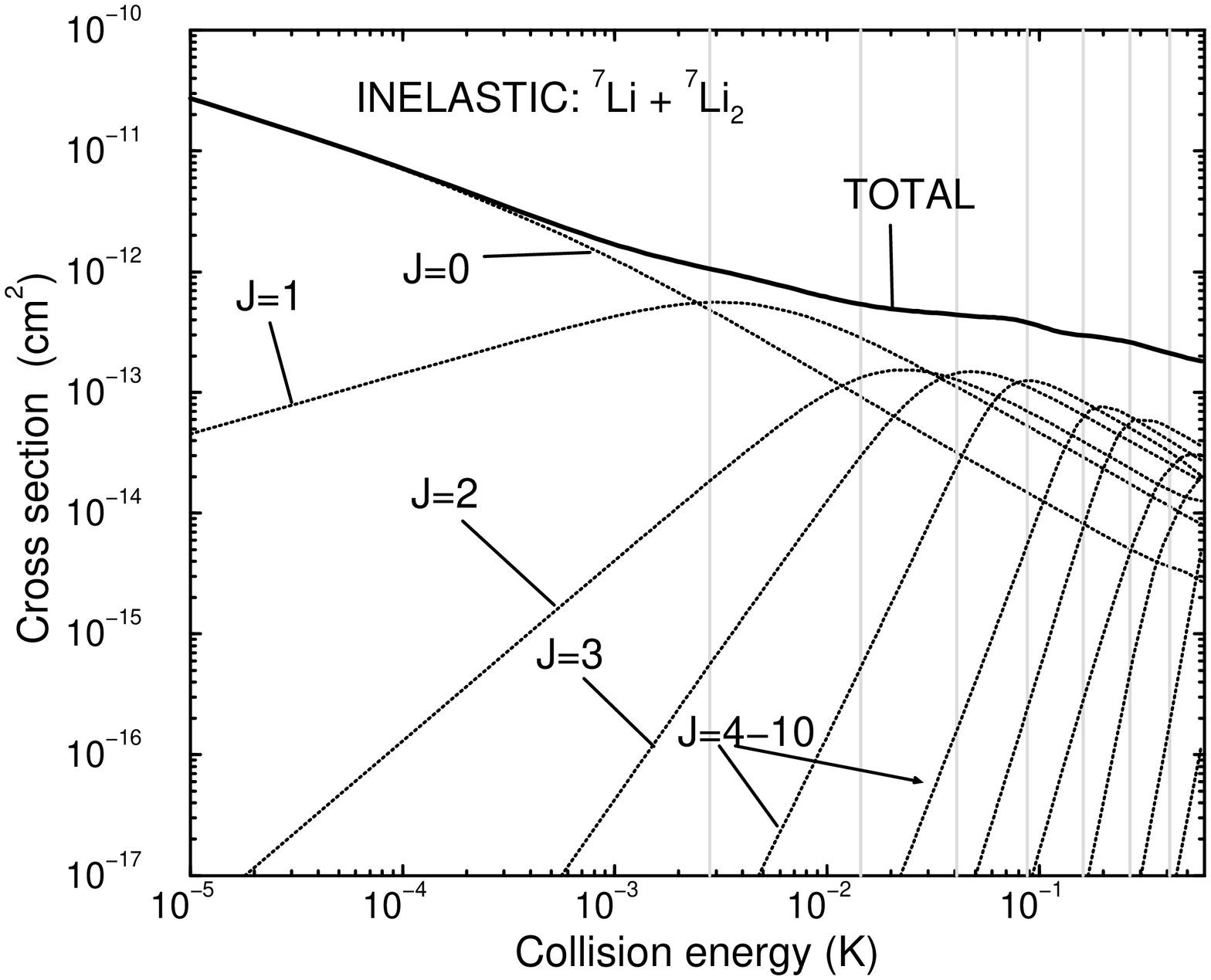}}}
\caption{Elastic cross sections (upper panel) and inelastic cross
sections (lower panel) for $^7$Li + $^7$Li$_2$($v_i=1$, $n_i=0$),
with contributions from individual partial waves. The vertical
lines indicate centrifugal barrier heights for $l\ge1$.}
\label{xsec-bosons}
\end{center}
\end{figure}

\begin{figure} [htbp]
\begin{center}
\rotatebox{270}{ \resizebox{7cm}{!}
{\includegraphics{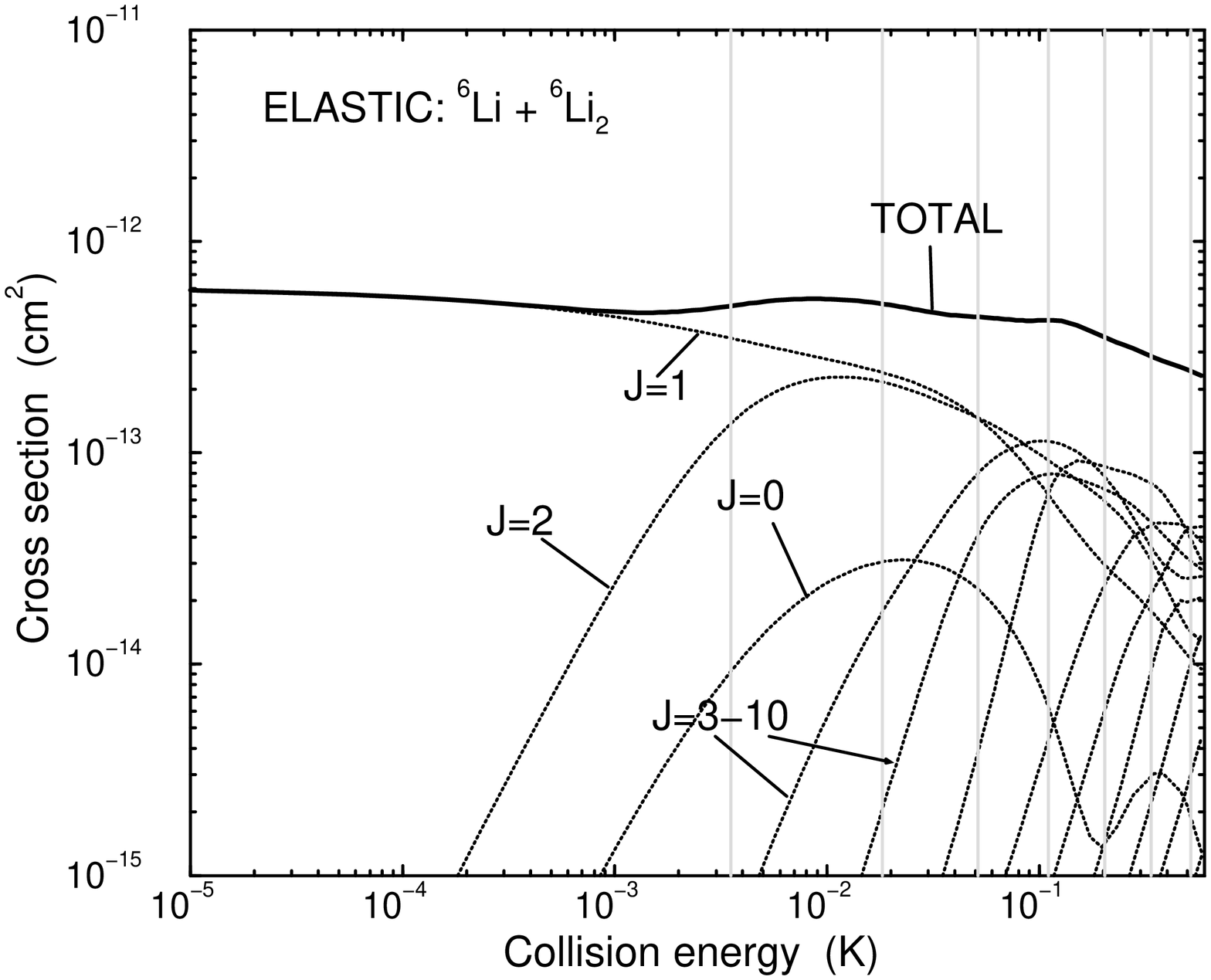}}} \rotatebox{270}{
\resizebox{7cm}{!} {\includegraphics{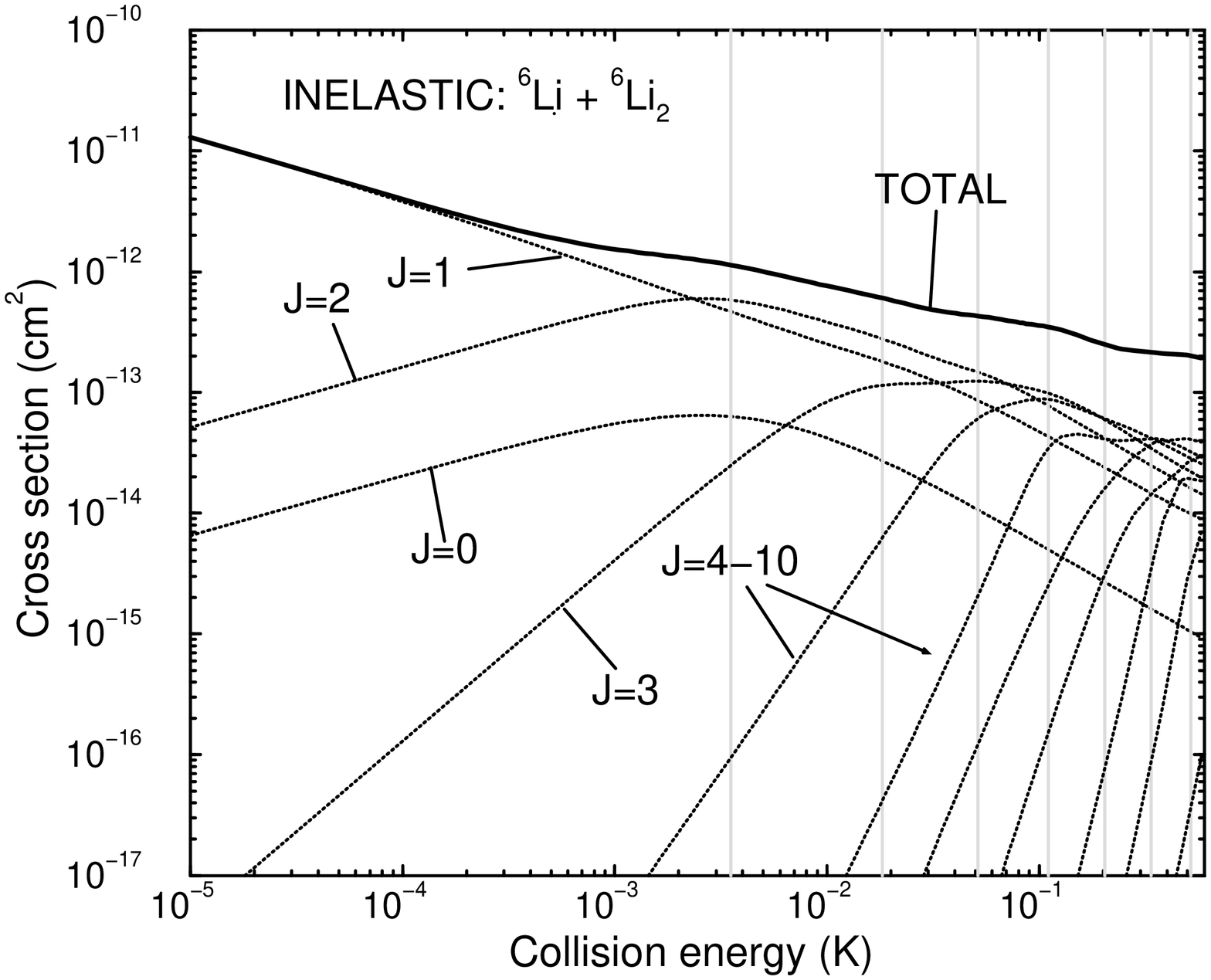}}}
\caption{Elastic cross sections (upper panel) and inelastic cross
sections (lower panel) for $^6$Li + $^6$Li$_2$($v_i=1$, $n_i=1$),
with contributions from individual partial waves. The vertical
lines indicate centrifugal barrier heights for $l\ge1$.}
\label{xsec-fermions}
\end{center}
\end{figure}

We have calculated the low-energy elastic and total quenching
cross sections $\sigma(E)$ for $^7$Li$_2$ colliding with $^7$Li
and for $^6$Li$_2$ colliding with $^6$Li, for bosons and fermions
in all allowed initial states $(v,n)$ with $v\le3$ and $n\le11$.
The cross sections for $v=1$ (with $n=0$ for $^7$Li$_2$ and $n=1$
for $^6$Li$_2$) are shown as a function of collision energy $E$ in
Figs.\ \ref{xsec-bosons} and \ref{xsec-fermions}, together with
the contributions from individual partial waves $J$. The partial
wave sums are converged up to 500 mK for $J_{\rm max}=10$. For
$v=1$, both the elastic and inelastic cross sections are slightly
smaller for $^6$Li than for $^7$Li at energies below 1 mK.
However, the difference is only about a factor of 2, and it is
{\it not} due to the difference between boson and fermion
symmetries: for other values of $v$ and $n$ the ratio is often
reversed.

At very low energies (up to about 1 mK), the cross sections are
dominated by collisions with orbital angular momentum $l=0$, for
which there are no centrifugal barriers. A consequence of the
symmetry constraints is that total angular momentum $J=0$ has an
$l=0$ channel for $^7$Li + $^7$Li$_2$ but not for $^6$Li +
$^6$Li$_2$; in the latter case the lowest partial wave containing
$l=0$ is $J=1^-$, where the superscript indicates odd parity. For
this reason, low-energy calculations on $^6$Li are significantly
more expensive than those on $^7$Li.

For nearly all initial $v$ and $n$, and for both bosons and
fermions, the ratio of inelastic to elastic cross sections is
greater than 30 at $E=1\ \mu$K and remains greater than 1
throughout the Wigner regime (i.e., up to about 1 mK). The only
exceptions to this are for rotational quenching of the
$(v,n)=(0,2)$ state for bosons (ratio 4.3 at 1 $\mu$K) and the
(0,3) state for fermions (ratio 7.5), for which there is only one
inelastic channel with $l=0$.

Scattering calculations for Li + Li$_2$ are converged up to about
10$^{-4}$ K using just a single partial wave ($J=0$ for $^7$Li and
$J=1^-$ for $^6$Li). At higher energies, other partial waves start
to contribute significantly. For $^7$Li the contributions decrease
monotonically with $J$ at low energies because of centrifugal
barriers corresponding to $l=J$ for $n=0$. For $^6$Li, the
situation is slightly different: each partial wave is made up of a
``parity favored'' block (with parity $(-1)^J$) and a ``parity
unfavored'' block; the former includes an $n=1$ channel with
$l=J-1$, while for the latter the lowest $l$ is $l=J$.

At collision energies below the centrifugal barrier, the partial
cross sections follow Wigner laws, $\sigma^{l}_{\rm inel} \sim
E^{l-1/2}$. Since $k_{\rm inel}(E) = v_{\rm coll}\sigma_{\rm
inel}(E)$, where $v_{\rm coll}=(2E/\mu)^{1/2}$ is the collision
velocity and $\mu$ is the atom-diatom reduced mass, the inelastic
rate coefficient is independent of $E$ at limitingly low energies.
For initial $v=1$, the results in Figs.\  \ref{xsec-bosons} and
\ref{xsec-fermions} give calculated limiting values $k_{\rm
inel}=5.6\times 10^{-10}$ cm$^3$ s$^{-1}$ for bosons and
$2.8\times 10^{-10}$ cm$^3$ s$^{-1}$ for fermions.

\begin{figure} [htbp]
\begin{center}
\rotatebox{270}{ \resizebox{7cm}{!}
{\includegraphics{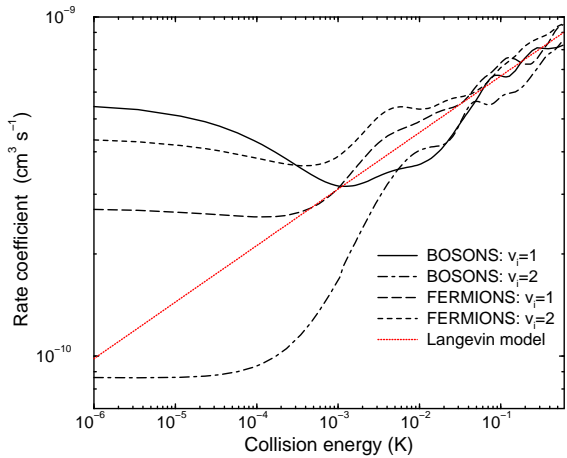}}} \caption{Total inelastic
rate coefficients for collisions of Li with Li$_2$ ($v=1$ and 2,
with $n=0$ for bosons and $n=1$ for fermions).} \label{capture}
\end{center}
\end{figure}

At energies above the barrier height, the inelastic probabilities
saturate at values close to 1 and the energy dependence in each
partial wave is then governed by the prefactor $E^{-1}$ that
multiplies the S-matrix term in the expression for the cross
section. At energies high enough that several partial waves
(typically $>3$) contribute significantly to the overall cross
sections, the energy dependence is described approximately by
classical Langevin capture theory \cite{levine},
\begin{equation}
k_{\rm inel}(E) = 3\pi (2E/\mu)^{1/2}
\left(\frac{C_6}{4E}\right)^{1/3}, \label{langevin}
\end{equation}
where $C_6=3086\ E_ha_0^6$ is the isotropic atom-diatom dispersion
coefficient. The Langevin result is compared with the full quantum
results for the quenching rates for $v=1$ and 2 in Fig.\
\ref{capture}; it may be seen that the Langevin and full quantum
results agree semi-quantitatively at collision energies above
about 10 mK for both bosons and fermions.

The atom--atom part of the potential used in our calculations
gives triplet scattering lengths that are approximately correct
for both $^7$Li ($-26.35$ $a_0$) and $^6$Li (Â$-1531$ $a_0$).
However, at the magnetic field used in the molecule production
experiments \cite{Grimm03a}, the spin states involved had
$a\approx +3500\ a_0$, and it is important to establish whether
fermionic suppression of quenching rates occurs for such large
positive atom-atom scattering lengths. We have therefore repeated
the calculations for energies $E=1\ \mu$K and 1 nK with the
short-range part of the 2-body potential adjusted to reproduce
$a\approx 3500\ a_0$. (Slightly different adjustments were needed
for bosons and fermions). Under these circumstances the calculated
quenching rates for initial $v=1$ at limitingly low energy are
$7.7\times 10^{-11}$ cm$^3$ s$^{-1}$ for $^7$Li and $2.9\times
10^{-10}$ cm$^3$ s$^{-1}$ for $^6$Li. This confirms that there is
no fermionic suppression for the low-lying states that we
consider.

The Langevin result, Eq.\ \ref{langevin}, may be used to predict
inelastic rates for other atom-diatom alkali systems in
spin-polarized states. Outside the Wigner regime, the inelastic
rate coefficients for different systems will be proportional to
$C_6^{1/3}/\mu^{1/2}$. The resulting inelastic rate coefficients
for $^{23}$Na, $^{40}$K, $^{87}$Rb and $^{133}$Cs are lower than
that for lithium by factors of 1.81, 1.75, 2.43, and 2.65,
respectively, using $C_6$ coefficients equal to twice the atomic
dispersion coefficients from Ref.\ \onlinecite{Der99}. The lower
bound of applicability of the model may be estimated from the
centrifugal barrier heights for $l=3$, which are 6.86 mK, 1.89 mK,
0.537 mK, and 0.235 mK, for Na, K, Rb, Cs, respectively.

In conclusion, we have carried out full quantum dynamics
calculations of vibrational relaxation (quenching) of $^7$Li$_2$
by $^7$Li and of $^6$Li$_2$ by $^6$Li, taking full account of the
boson and fermion symmetries. We find that for low initial
vibrational states of the molecules ($v\le 3$), there is {\it no}
systematic suppression of the quenching rates for molecules formed
from fermionic atoms, even when the atom-atom scattering length is
large and positive. This contrasts with the situation for
molecules formed in the highest vibrationally excited state
\cite{Grimm03a,Jin03a, Kett03a}.

Collisions of molecules involving {\it mixed} isotopes offer
fascinating new possibilities and in some cases allow identifiable
chemical reactions even between ground-state molecules. For
example, the reaction $^7$Li + $^6$Li$_2(v=0)$ $\rightarrow$
$^6$Li + $^6$Li$^7$Li$(v=0$) is exothermic by 2.643 K because of
the difference in zero-point energy of the two diatomic molecules.
Calculations on collisions such as this will be described in
future work.

\acknowledgments

PS and JMH are grateful to EPSRC for support under research grant
no.\ GR/R17522/01. MTC is grateful for sponsorship from the
University of Durham and for an ORS award from Universities UK.

\end{document}